# To Use or Not to Use Diagrams: The Effect of Drawing a Diagram in Solving Introductory Physics Problems

Alexandru Maries and Chandralekha Singh

*Department of Physics and Astronomy, University of Pittsburgh, Pittsburgh, PA 15260*

**Abstract.** Drawing appropriate diagrams is a useful problem solving heuristic that can transform a given problem into a representation that is easier to exploit for solving it. A major focus while helping introductory physics students learn problem solving is to help them appreciate that drawing diagrams facilitates problem solution. We conducted an investigation in which 111 students in an algebra-based introductory physics course were subjected to two different interventions during recitation quizzes throughout the semester. They were either (1) asked to solve problems in which the diagrams were drawn for them or (2) explicitly told to draw a diagram. A comparison group was not given any instruction regarding diagrams. We developed a rubric to score the problem-solving performance of students in different intervention groups. We investigated two problems involving electric field and electric force and found that students who draw expert-like diagrams are more successful problem solvers and that a higher level of detail in a student's diagram corresponds to a better score.

## INTRODUCTION

Drawing diagrams is a useful problem solving heuristic. Diagrammatic representations have been shown to be superior to exclusively employing verbal representations when solving problems [1,2]. This is one reason why physics experts automatically employ diagrams in attempting to solve problems. However, introductory physics students need explicit help understanding that drawing a diagram is an important step in organizing and simplifying the given information into a representation which is more suitable to mathematical manipulation. Previous research shows that students who draw diagrams even if they are not rewarded for it are more successful problem solvers [3]. Investigations into how drawing free body diagrams (FBDs) affects problem solving show that only drawing correct FBDs improves a student's score and that students who draw incorrect FBDs do not perform better than students who draw no diagrams [4] and that prompting students to draw FBDs in introductory mechanics results in students attempting to follow formally taught problem solving methods rather than intuitive methods and deteriorated performance [5]. We extend that research here and investigate how the quality of a diagram affects performance. We also compare performance on 1D and 2D electric force problems.

## METHODOLOGY

A class of 111 algebra-based introductory physics students was broken up into three different recitations. All recitations were taught in the traditional way in which the TA worked out problems similar to the homework problems and then gave a 15 minute quiz at the end of class. Students in all recitations attended the same lectures, were assigned the same homework, and had the same exams and quizzes. In the recitation quizzes throughout the semester, the three groups were given the same problems but with the following interventions: in each quiz problem, the first intervention group, which we refer to as "prompt only group" or "PO", was given explicit instructions to draw a diagram with the problem statement; the second intervention group (referred to as "diagram only group" or "DO") was given a diagram drawn by the instructor that was meant to aid in solving the problem and the third group was the comparison group and was not given any diagram or explicit instruction to draw a diagram with the problem statement ("no support group" or "NS").

The sizes of the different recitation groups varied from 22 to 55 students because the students were not assigned a particular recitation; they could go to whichever recitation they wanted. For the same reason, the sizes of each recitation group also varied from week to week, although not as drastically because most students ($\approx 80\%$) would stick with a particular recitation. Furthermore, each intervention was not matched to a particular recitation. For example, in one week, NS was the Tuesday recitation while another week NS was a different recitation section. This is important because it implies that individual students were subjected to different interventions from week to week and we do not expect cumulative effects due to the same group of students always being subjected to the same intervention.

In order to ensure homogeneity of grading, we developed rubrics for each problem we analyzed and made sure that there was at least 90% inter-rater-reliability between two different raters. The development of the rubric for each problem went through an iterative process. During the development of the rubric, the two graders also discussed a student's score separately from the one obtained using the rubric and adjusted the rubric if it was agreed that the version of the rubric was too stringent or too generous. After each adjustment, all students were graded again on the improved rubric.

We analyzed two problems; the first problem is one dimensional and has two almost identical parts, one on electric field and the other on electric force. The second problem is a two dimensional problem on electric force. The two problems and the diagrams (given only to students in DO) are the following:

**Problem 1**

"Two equal and opposite charges with magnitude $10^{-7}$ C are held 15 cm apart.
(a) What is the magnitude and direction of the electric field at the point midway between the charges?
(b) What is the magnitude and direction of the force that would act on a $10^{-6}$ C charge if it is placed at that midpoint?"

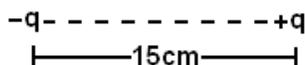

**FIGURE 1.** Diagram for Problem 1 given only to students in DO.

**Problem 2**

"Three charges are located at the vertices of an equilateral triangle that is 1 m on a side. Two of the charges are 2 C each and the third charge is 1 C. Find the magnitude and direction of the net electrostatic force on the 1 C charge."

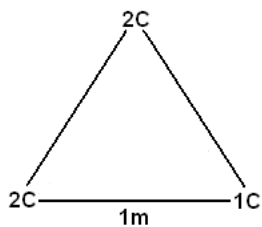

**FIGURE 2.** Diagram for Problem 2 given only to students in DO.

These diagrams were drawn by the instructor and they are very similar to what most physics experts would draw initially in order to solve the problem. Neither diagram was meant to trick the students, but rather they were provided as a scaffolding support.

We developed rubrics for each problem. For Problem 1, we were interested in comparing student performance on electric field with electric force. Therefore, parts (a) and (b) were scored separately. In Table 1, we provide the summary of the rubric for part (a) (electric field) of the first problem. The rubric for part (b) (electric force) is similar.

Table 1 shows that there are two parts to the rubric: Ideal Knowledge and Incorrect Ideas. Table 1 also shows that in the Ideal Knowledge part, the problem was divided into different sections and points were assigned to each section. Each student starts out with 10 points and in the Incorrect Ideas part we list the common mistakes students made and how many points were deducted for each of those mistakes. Each mistake is connected to a particular section (mistake labeled 1 is for section 1, mistakes labeled 2.1 through 2.5 are for section 2 etc.) and it is not possible to deduct more points than a section has (the mistakes labeled 2.1 and 2.2 are exclusive with all other mistakes in section 2 and with each other). We also left ourselves a small window (labeled 2.5) if the mistake a student made was not explicitly in the rubric.

**TABLE 1.** Summary of the rubric for part (a) of Problem 1 ("E" stands for electric field).

| Ideal Knowledge | |
|---|---|
| 1. Used correct equation for E | 1p |
| 2. Added the two fields due to individual charges correctly | 7p |
| 3. Direction for net electric field | 1p |
| 4. Units | 1p |
| **Incorrect Ideas** | |
| 1. Used wrong equation for E (-1p) | |
| 2.1 Did nothing in this section (-7p) | |
| 2.2 Did not find electric fields due to both charges (-6p) | |
| 2.3 Used Pythagorean theorem (not relevant here) or obtained zero for electric field (-4p) | |
| 2.4 Did not use half distance for E (-2p) | |
| 2.5 Minor mistake(s) in finding E (-1p) | |
| 3. Wrong or no mention of direction of net electric field (-1p) | |
| 4. Wrong or no units (-1p) | |

## RESULTS – Problem 1

We investigated differences between the intervention groups resulting from the differences in instructions regarding diagrams (draw a diagram, diagram given, or no instructions). We found that although all the students had a diagram drawn for this problem (whether they drew it themselves or had it drawn for them) regardless of the instructions they received, those prompted to draw a diagram (PO) were more likely to draw expert-like diagrams. We considered that an expert-like diagram should have, in addition to the two charges, either two electric field or two electric force vectors or all four explicitly drawn at the midpoint. Any other diagram was considered to not be expert-like (for example, a diagram containing just the two charges or diagrams containing the two charges and arrows drawn somewhere other than at the midpoint). It is worthwhile to note that students in DO were given a diagram containing the two charges (not expert-like). We thought that some students might modify it by adding vectors at the midpoint that indicate the direction of electric fields or electric forces in order to make it expert-like. Therefore, in addition to investigating the number of students who drew expert-like diagrams in each intervention group, we also investigated the number of students in each group who had diagrams of only the two charges. The results are shown in Table 2.

**TABLE 2.** Percentages (and numbers) of students who drew expert-like diagrams ("Expert d.") and who only drew two charges ("Only 2 charges") in each intervention group.

| Quiz | Expert d. | Only 2 charges |
|---|---|---|
| PO | 66% (19) | 14% (4) |
| DO | 45% (18) | 48% (19) |
| NS | 41% (21) | 33% (17) |

Table 3 lists p values for comparison between data in Table 2; e.g., the value 0.036 under "PO-NS" for "Expert d." means the p value for comparison of the percentage of students who used expert-like diagrams in PO and NS is 0.036.

**TABLE 3.** The p values for comparison of percentage of students who drew expert-like diagrams ("Expert d.".) and who drew only the two charges ("Only 2 ch.") in the different intervention groups.

|  | PO-DO | PO-NS | DO-NS |
|---|---|---|---|
| Expert d. | 0.092 | 0.036 | 0.190 |
| Only 2 ch. | 0.001 | 0.056 | 0.170 |

Table 3 shows that students in PO are statistically more likely to draw expert-like diagrams than students in NS. It also shows that students in DO are statistically more likely than students in PO to use a diagram of only the two charges. Students in DO were given this diagram so if they were using this type of diagram, they had not modified it in order to make it expert-like (Table 2 shows that almost half of them did not modify it to make the diagram provided expert-like). On the other hand, Table 2 shows that students in PO who were asked to draw a diagram were more likely to draw and use expert-like diagrams. Below, we provide evidence that for Problem 1, drawing an expert-like diagram improves students' scores.

We stratified all the students in three categories based on the quality of their diagrams and analyzed their scores as shown in Table 4. A lower category corresponds to a lower quality diagram. $T$-tests on the data in Table 4 reveal that students who drew expert-like diagrams (Categories 2 and 3) performed better than students who drew other diagrams (Category 1). Both p values were small ($<0.001$) showing statistically significant differences.

**TABLE 4.** Sizes (N), averages and standard deviations for groups of students with different quality diagrams. Category 1 consists of students who drew a diagram that was not expert-like, Category 2 consists of students who drew either two electric field or two electric force vectors at the midpoint and Category 3 consists of students who drew all four vectors at the midpoint.

| Quiz | N | Avg. | Std.dev. |
|---|---|---|---|
| Category 1 | 62 | 6.4 | 2.58 |
| Category 2 | 49 | 8.3 | 2.20 |
| Category 3 | 9 | 8.9 | 1.36 |

## RESULTS - Problem 2

For Problem 2, a two dimensional (2D) problem, there were no statistically significant differences between the different intervention groups, both in terms of scores and in terms of percentages of students drawing expert-like diagrams. One possible explanation for this result is that Problem 2 is two dimensional and it is very difficult (perhaps even impossible) to solve by a novice without the use of an expert-like diagram which would at least include the directions of the two electric forces acting on the 1C charge. Therefore, students in all intervention groups were more likely to draw expert-like diagrams regardless of the instructions they received involving diagrams. However, we found that there was a relationship between the level of detail in a student's diagram and their performance.

Similar to what we did for Problem 1, we stratified the students based on three categories of diagram quality (or detail). Category 1 corresponds to diagrams with just the three charges, Category 2 corresponds to diagrams with the three charges and the two forces acting on the 1C charge and Category 3 corresponds to diagrams with the three charges, the two forces acting on the 1C charge, and the $x$ and $y$ components of those forces. Since students were explicitly asked to indicate the direction of the net force acting on the 1C charge, whether or not a student drew a vector for the net force was not taken into consideration when choosing the categories. We only took into consideration levels of detail that the students themselves thought would help them solve the problem, not what they were explicitly asked to draw. The results are shown in Table 5.

Table 5 shows a relationship between the level of detail in the diagrams drawn and the score. In particular, a higher level of detail (represented here by a higher category number) corresponds to a better score. *T*-tests on the data in Table 5 shows that students who, in addition to the three charges, drew two force vectors (Category 2), outperformed those who only drew the three charges (Category 1) ($p = 0.008$). Similarly, those who drew the two forces due to individual charges and their *x* and *y* components (Category 3), outperformed those who drew only the two forces (Category 2) ($p < 0.001$).

**TABLE 5.** Sizes (N), averages and standard deviations for students in different categories (by diagram detail).

|  | N | Avg. | Std.dev. |
|---|---|---|---|
| Category 1 | 27 | 4.1 | 2.49 |
| Category 2 | 58 | 5.7 | 2.92 |
| Category 3 | 33 | 8.0 | 2.22 |

## DISCUSSION

We found that for Problem 1, students who were explicitly prompted to draw a diagram were more likely to draw expert-like diagrams. We also found that students who drew expert-like diagrams performed better than students who drew other kinds of diagrams. Among the students in DO who were provided with a diagram (not expert-like unless modified by the student by adding force and/or field arrows) less than half attempted to draw the arrows, which is statistically significantly lower than those who were not provided any diagram and explicitly asked to draw it (NS). This finding suggests that in an algebra-based introductory physics course prompting students to draw diagrams is more likely to provide better scaffolding support than giving students a basic diagram and perhaps this instruction should be incorporated in quiz/midterm problems for helping students learn effective problem solving strategies.

We also found that more detailed diagrams (in general a more detailed diagram is also a higher quality diagram) correspond to better performance. Unlike Rosengrant's study [4], in our study here, the correctness of the diagrams (correctness of the vector arrows representing electric fields or forces) did not impact students' performance significantly. The reason is that for both problems in this study, students who drew incorrect vector arrows in the diagrams had the wrong direction of electric field or electric force vectors due to both charges. These students differ from those with the correct direction in that they obtained a direction for net electric field or force which was 180º from the correct direction. This mistake would cost a student 1 out of 10 points because partial credit was given. On the other hand, Ref [4] involved multiple choice problems only and since sometimes alternative choices to the multiple choice problems were based on common student errors, one would expect that the correctness of the diagrams would make a larger difference in Rosengrant's study than in ours.

One theoretical model that can provide a possible explanation for why students with more detailed diagrams performed better is the cognitive load theory which incorporates the notion of distributed cognition [6]. In Problem 2, students had to add forces by using components, so those who did not even draw the force vectors they needed to add vectorially would have to keep too much information in their working memory while engaged in problem solving (individual components of the two forces, angles required to obtain those components, trigonometric functions needed for each component etc.). This can lead to cognitive overload and deteriorated performance. Explicitly drawing the forces and their components can reduce the amount of information in the working memory at a given time while engaged in problem solving and may therefore make the student more able to go through all the steps necessary without making mistakes.

It is also important to note that these problems were given in the second semester of a one year introductory physics course for algebra based students. These students had done problems where they had to find the net force in Newtonian mechanics, and still less than 30% realized that they should draw the components of the electric force in the 2D problem presented here. Also, only 42% of all the students indicated a direction for the net force. This is an indication of a lack of transfer from one context to another. Students' performance also suggests that most introductory students do not employ good problem solving heuristics and their familiarity with addition of vectors may also require an explicit review. Earlier surveys have found that only about 1/3 of the students in an introductory physics class have enough knowledge about vectors to begin the study of Newtonian mechanics [7]. Here we find that even after a semester of instruction in physics that involves a significant amount of vector addition, the fraction remains about the same and students still have difficulty dealing with vector addition in component form.

## REFERENCES

1. J. Larkin and H. Simon, *Cog. Sci.* **11**, 65-99 (1987).
2. J. Zhang and D. Norman, *Cog. Sci.* **18**, 87-122 (1994).
3. A. Mason and C. Singh, *Am. J. Phys.* **78(7)**, 748-754, (2010).
4. D. Rosengrant, Ph.D. Dissertation, Rutgers University (2007).
5. A. Heckler, *Int. J. Sci. Ed.* **14**, 1829-1851 (2010).
6. A. Newell and H. Simon, *Human Problem Solving,* Englewood Cliffs NJ: Prentice Hall, 1972.
7. R. K. Knight, *The Physics Teacher* **33,** 74-80 (1995).